\documentclass[sigplan,nonacm]{acmart}
\usepackage[utf8]{inputenc}
\usepackage{natbib}
\usepackage[noframes,novert]{ffcode}
\usepackage{afterpage}
\usepackage{to-be-determined}
\usepackage{float}
\usepackage{href-ul}
\usepackage{paralist}
\usepackage{caption}
\usepackage{soul}
\usepackage[capitalize]{cleveref}
\usepackage{booktabs}
\usepackage{tabularx}

\title{Java Classes with ``-Er'' and ``-Utils'' Suffixes Have Higher Complexity}

\author{Anna Sukhova}
\orcid{0009-0000-8931-8813}
\email{agsukhova_1@edu.hse.ru}
\affiliation{\institution{HSE}\city{Moscow}\country{Russia}}

\author{Alexey Akhundov}
\orcid{0009-0003-8218-0710}
\email{anakhundov@edu.hse.ru}
\affiliation{\institution{HSE}\city{Moscow}\country{Russia}}

\author{Efim Verzakov}
\email{eaverzakov@edu.hse.ru}
\orcid{0009-0006-7459-5782}
\affiliation{\institution{HSE}\city{Moscow}\country{Russia}}

\author{Yegor Bugayenko}
\orcid{0000-0001-6370-0678}
\email{yegor256@gmail.com}
\affiliation{Huawei\city{Moscow}\country{Russia}}


\newcommand\eror{``\ff{-Er}/\ff{-Or}''}
\newcommand\utils{``\ff{-Utils}''}

\begin{abstract}
In object-oriented programming languages, a belief exists that classes with \eror{} and \utils{} suffixes are ``code smells'' because they take over a lot of functional responsibility, turning out to be bulky and complicated, and therefore making it more difficult to maintain the code. In order to validate this intuition, we analyzed complexity and cohesion of 13,861 Java classes from 212 unique open-source GitHub repositories. We found out that average values of Cyclomatic Complexity and Cognitive Complexity metrics are at least 2.5 times higher when suffixes are present.
\end{abstract}

\begin{document}

\maketitle

\section{Introduction}

Some object-oriented practitioners believe that the introduction of classes with names ending in \eror{} or \utils{}, later called ``functors''~\citep{cunningham2014}, is a bad design decision~\citep{code-quality-study,reusable-components} due to the seemingly increasing level of code understandability and the decreasing level of code maintainability. They argue that functors can be less cohesive~\citep{stevens1974structured} and more complex because they tend to provide common methods for diverse purposes, which are reused by other classes~\citep{selby1991analyzing}, so the methods are weakly connected with each other~\citep{bug-1,jehad-1}.

Programmers often frame complex or repetitive procedures, such as file operations, encoding/decoding, or network interactions, in so called ``Utility Classes''~\citep{hamou2004reasoning,gueheneuc2004systematic}. To name an example, Apache Commons\footnote{\url{https://commons.apache.org/}}, a widely spread and used series of Java libraries, have a significant portion of their classes named with the \utils{} suffix. These classes are frequently criticized for violating the object-oriented design principles~\citep{beecher2018classes} and making design less structured~\citep{selby1991analyzing}.

We also presume that functors, which typically execute operations on behalf of other classes and amalgamate numerous responsibilities, lead to code that is more complex and less cohesive~\citep{stevens1974structured}. However, these assertions lack empirical support and are often based on personal experience. To validate our hypothesis, we analyzed 13,861 Java classes from 1,000 unique open-source GitHub repositories and discovered that, on average, the complexity of methods in functor classes is more than 2.5 times higher than in other classes. We didn't find, though, that functor classes are less cohesive. Despite the relatively small dataset under examination, we conclude that functor classes represent poor design.

This article is structured as follows:
\cref{sec:related} presents a review of works related to the research,
\cref{sec:method} outlines practical steps taken during the study,
\cref{sec:results} describes obtained results,
\cref{sec:discussion}
 provides interpretations of our findings,
\cref{sec:limitations} explores limitations, and
\cref{sec:conclusion} offers a summary of the paper.

\section{Related Work}
\label{sec:related}

Among others, \citet{eo1,jehad-1,cunningham2014} have emphasized that functor classes pose a problem for maintainability, as they typically serve as mere aggregates of procedures that are barely related to each other. \citet{bugayenko2014blog0505} even argued that functors are not ``true'' objects and do not belong in the domain of object-oriented programming~\citep{west2004object} because they do not encapsulate anything and thus have no state, which, according to \citet{grady2007object}, is a paramount quality of an object.

\citet{oizumi-1} described a technique used for identifying potential design-related issues in software and illustrated the example of a design problem, termed ``Fat Interface,'' which describes a situation where an interface exposes multiple functionalities through a general interface. This example includes \eror{} classes as well as their inter-relationships. The paper includes further elaboration on such code anomalies as Long Method and God Class.

\citet{griggs2011} highlighted another problem related to \eror{} classes: such classes concentrate on behavior rather than data that they store. They argue that this is a violation of object-oriented paradigm: ``We want to create objects that describe what they are, and then bind behavior to them.'' They even proposed a re-naming convention for some patterns, which may help avoid functor classes, e.g. ``\ff{-Registry}'' instead of ``\ff{-Manager}'' and ``\ff{-Analysis}'' instead of ``\ff{-Analyzer}''.

\citet{pescio2011} pointed that a class name keeps functions together and describes what code is truly doing. They explained for bad naming conventions (the ``\ff{-Er}'' suffix, the ``\ff{-Able}'' suffix, the ``\ff{-Object}'' suffix, and the ``\ff{I-}'' prefix) and their influence on bad properties of objects. For example, ``\ff{-Manager}'' and ``\ff{-Managed}'' classes are procedures doing the real work and plain old data structure, these classes do not represent real objects. Also, they discovered the interface ``\ff{IValidatableObject}`` from the .NET Framework 4.0, which  ``provides a way for an object to be invalidated'' and ``contains three of four bad class naming practices.''

However, none of the articles arguing against functor classes had any metrical proof of how ``bad'' names of classes are correlated with their complexity or cohesion.

\section{Method}
\label{sec:method}

The goal of this study was to statistically verify that functors are bad design decisions due to decrease of cohesion and increase of complexity by analyzing large number of open-source projects and evaluating code quality metrics. We were interested in answering the following research questions:
\begin{description}
    \item[RQ1] Do functors have lower cohesion?
    \item[RQ2] Do functors have higher complexity?
\end{description}

First, we took an existing ``Classes and Metrics'' (CAM) dataset~\citep{Bugayenko_CAM_A_Collection_2024}, where Java classes were already collected from open-source repositories and more than 40 metrics were already calculated. We chose CAM because it enabled us to collect open-source data conveniently and provided pre-existing metrics.

Our preference for open-source data was based on the diversity of open-source projects, the variety of tasks they handle, and their wide range of applications. We also aimed to make it easy for other researchers to replicate our work. Additionally, we didn't consider closed-source projects, either due to limitations of access or a lack of research permissions.

Then, we split all Java classes into three groups:
\begin{inparaenum}[1)]
\item classes with the \utils{} suffix,
\item classes with \eror{} suffixes,
and
\item all other classes.
\end{inparaenum}
To the first group we also put classes with the following suffixes: ``\ff{-Utils},'' ``\ff{-Util},'' ``\ff{-Utilities},'' and ``\ff{-Utility}.'' Additionally, we filtered out classes from the second group with subsequent suffixes:

\begin{ffcode}
Inner, Actor, Logger, Member, Order, Parameter,
Error, Calculator, Vector, Computer, Customer,
Trigger, Cluster, Cipher, Cursor, Number, Owner,
Meter, Letter, Answer, Author, Folder, Other,
Cashier, Broker, Motor, Mirror, Spider, Color,
Center, Layer, Never, Browser, Either, Tensor,
Cylinder, Meteor, Flower, Banner, Chapter,
Developer.
\end{ffcode}

Then, from the dataset we took the following metrics available for each Java class:
LCOM5~\citep{HendersonSellers1996CouplingAC},
Normalized Hamming Distance (NHD)~\citep{counsell},
Total Cyclomatic Complexity (CC)~\citep{McCabe1976ACM} of all methods in a class,
Total, Mimimum, Maximum and Average Cognitive Complexity (CoCo)~\citep{Campbell2018CognitiveC} of all methods in a class.

LCOM5 metric displays classes cohesion, more formally, the lack of it, by focusing on related methods and variables.
By definition~\citep{HendersonSellers1996CouplingAC}, LCOM5 is calculated for a class as following:
\begin{equation*}
\text{LCOM5} = \dfrac{a - k \cdot l}{l - k \cdot l}
\end{equation*}
where \(l\) is the number of class attributes, \(k\) is the number of methods and \(a\) is the sum of the number of distinct attributes that are accessed by each method in a class. Thus, the higher the value of metric the lower the cohesion, which indicates that introducing the evaluated class is considered to be a bad design decision.

NHD metric shows classes cohesion by using types of method's attributes, using the following formula~\citep{counsell}:
\begin{equation*}
\text{NHD} = 1 - \dfrac{2}{l k (k - 1)} \sum_{j=1}^l x_j (k - x_j)
\end{equation*}
where \(l\) is the number of distinct parameter types, \(k\) is the number of methods, \(a\) is the sum of the numbers of distinct parameter types of each method in the class and \(x_j\) is the the number of methods that have a parameter of type j. In contrast to LCOM5, this metric reaches the highest cohesion at its maximum.

CC metric is aimed at quantifying a program's complexity and understandability by counting its linearly independent paths, using the following formula~\citep{McCabe1976ACM}:
\begin{equation*}
C = E - N + 2
\end{equation*}
where \(E\) is the number of edges of the graph representation of code, \(N\) is the number of nodes in the graph.

CoCo remedies CC's shortcomings by punishing deviations of linear control flow which normally goes from top to bottom but not from left to right~\citep{Campbell2018CognitiveC}. It adds penalty for nesting more code: for example, adding loops in loops, maintenance of which requires more efforts. We examined aggregated values of CoCo for all methods in each class, such as aggregated total, average, minimum and maximum values.

Then, we filtered outliers---classes, for which the difference between the lines of code and number of blank lines was beyond thresholds of 0.01- and 0.99-quantilies amongst the whole data.

Then, we divided resulting data by the considered groups of classes as follows: classes with the \utils{} suffix, classes with \eror{} suffixes and rest of classes. For the last group we selected classes that do not contain static methods and attributes, because it could also be considered bad practice and makes it harder to distinguish between 'normal' classes and \utils{} classes, which usually contain static methods.

Then, to answer \textbf{RQ1}, we investigated the difference in cohesion metrics, specifically any increases or decreases, for \utils{} and \eror{} classes.

Then, in responding to \textbf{RQ2}, we observed the disparity in complexity metrics, particularly any upward or downward shifts, for classes with \utils{} and \eror{}.

During research, we did not account for constructors and did not consider them in calculations of metrics, because they are implicitly connected with class's methods due to initialization of attributes and therefore affect class cohesion as it is shown by \citet{biemann}. This connection also was demonstrated by \citet{bug-1}.

\section{Results}
\label{sec:results}

First, we took an existing dataset ``2023-10-22'' from the CAM framework~\citep{Bugayenko_CAM_A_Collection_2024}. This dataset comprises 863,000 Java classes out of 1000 open-source projects with associated metrics. We had to filter out the classes that didn't have any of the six metrics required for the research. We also removed outliers---classes, for which the difference between the lines of code and number of blank lines was beyond thresholds of 0.01- and 0.99-quantiles amongst the whole data. In the end, 13,861 classes were left in the dataset, classified into three groups, as shown in \cref{tab:size}: there are 5,610 classes with \eror{} suffixes, 72 classes with the \utils{} suffix, and 8,179 classes with no specific suffix.

After performing calculations described in \cref{sec:method}, we received metrics' values for three groups of interest corresponding to those with the prefix \eror{}, those with the suffix \utils{}, and the rest of the classes. The obtained values for classes are shown in \cref{tab:cohesion,tab:complexity}, where metrics are columns and classes are rows. The results show that functor classes, on average, are less cohesive and more complex. The values of both complexity metrics (CC and CoCo) for functor classes are almost three times larger than for other classes.

\begin{table}
\begin{tabular}{lrr}
\toprule
~ & LCOM5 & NHD \\
\midrule
\eror{} & \ul{0.835} & 0.533 \\
\utils{} & 0.810 & \ul{0.566}  \\
Rest & 0.704 & 0.562 \\
\bottomrule
\end{tabular}
\caption{The values of cohesion metrics for all three groups of Java classes, where LCOM5 and NHD refer to the corresponding cohesion metrics. The worst cases with the lowest cohesion are underscored (higher values of LCOM5 mean lower cohesion).}
\label{tab:cohesion}
\end{table}

\newcommand*\rot{\rotatebox{90}}
\begin{table}
\begin{tabular}{lrrrrr}
\toprule
~
    & \rot{CC}
    & \rot{CoCo}
    & \rot{ACoCO}
    & \rot{MxCoCo}
    & \rot{MnCoCo} \\
\midrule
\eror{}
    & 14.26
    & 20.106
    & \ul{3.790}
    & \ul{9.327}
    & \ul{1.867} \\
\utils{}
    & \ul{15.444}
    & \ul{20.931}
    & 3.514
    & 8.194
    & 1.583  \\
Rest
    & 5.983
    & 7.731
    & 1.876
    & 3.627
    & 1.218 \\
\bottomrule
\end{tabular}
\caption{The values of code complexity metrics for all three groups of Java classes, where ACoCo refers to the average CoCo of all methods in the class and MxCoCo/MnCoCo refer to the maximum/minimum values of all methods. The values of the highest complexity are underscored.}
\label{tab:complexity}
\end{table}

\begin{table}
\begin{tabular}{lrrr}
\toprule
~ & Classes & LoC & L/C \\
\midrule
\eror{} & 5,610 & 743337 & 132.502 \\
\utils{} & 72 & 9959 & 138.319  \\
Rest & 8,179 & 590892 & 72.245 \\
\bottomrule
\end{tabular}
\caption{Sizes of all three groups of Java classes under analysis, where ``Classes'' is the total count of Java classes that belong to the group, ``LoC'' is the total number of lines of code in all classes together, and ``L/C'' is a ratio of LoC per class.}
\label{tab:size}
\end{table}

\begin{figure}
\centerline{\includegraphics[width=.9\columnwidth]{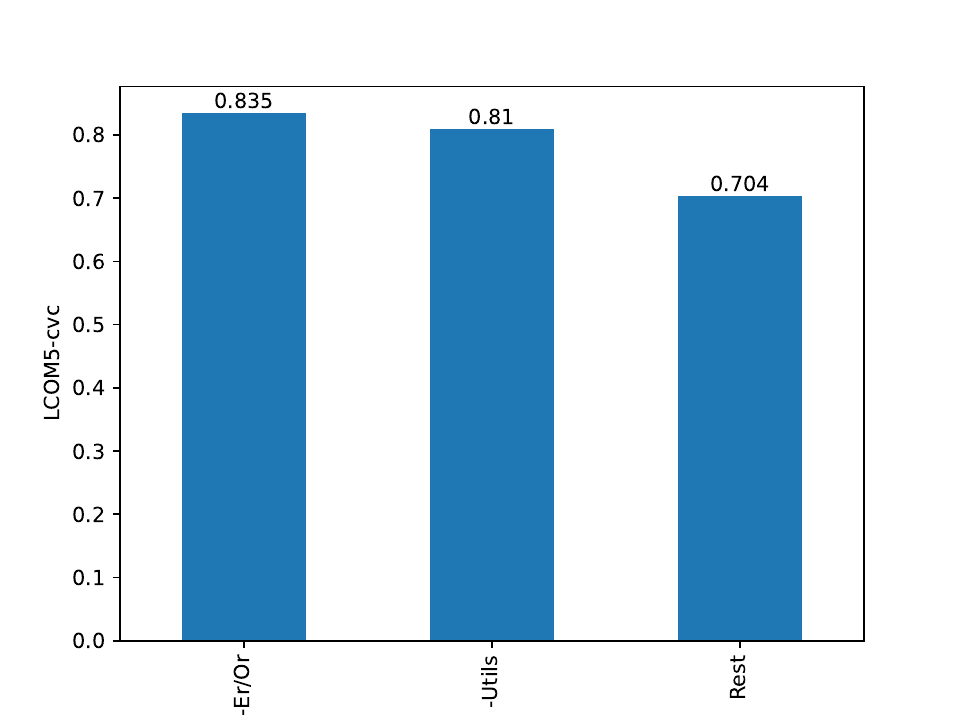}}
\caption{The average values of total LCOM5 for three groups of Java classes, not taking into account class constructors.}
\label{fig:lcom}
\end{figure}

\begin{figure}
\includegraphics[width=.9\columnwidth]{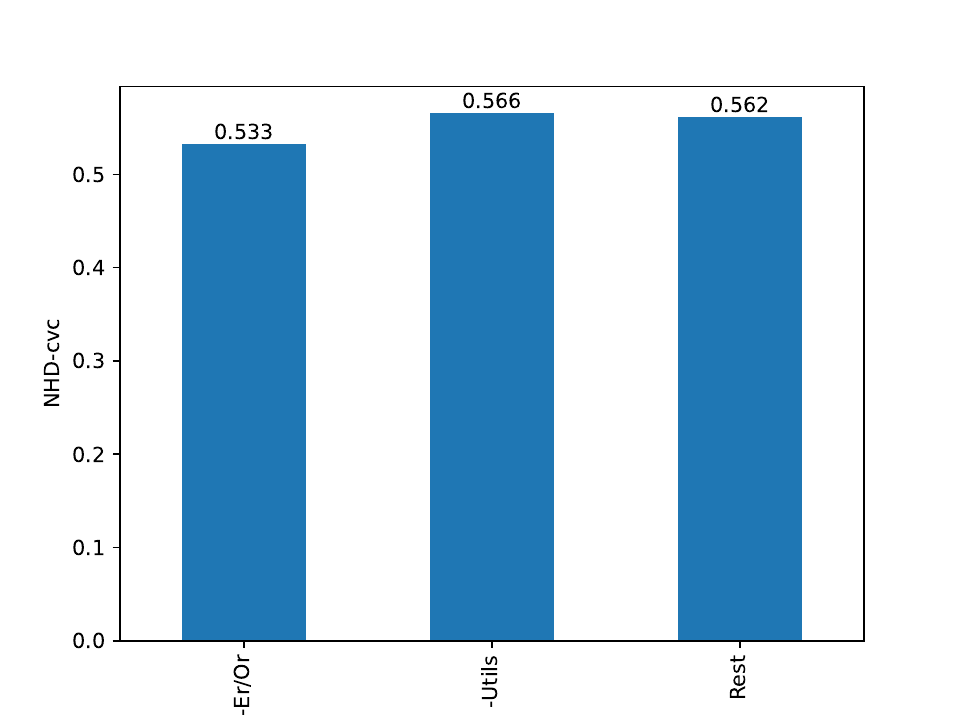}
\caption{The average values of total NHD for three groups of Java classes, not taking into account class constructors.}
\label{fig:nhd}
\end{figure}

\begin{figure}
\includegraphics[width=.9\columnwidth]{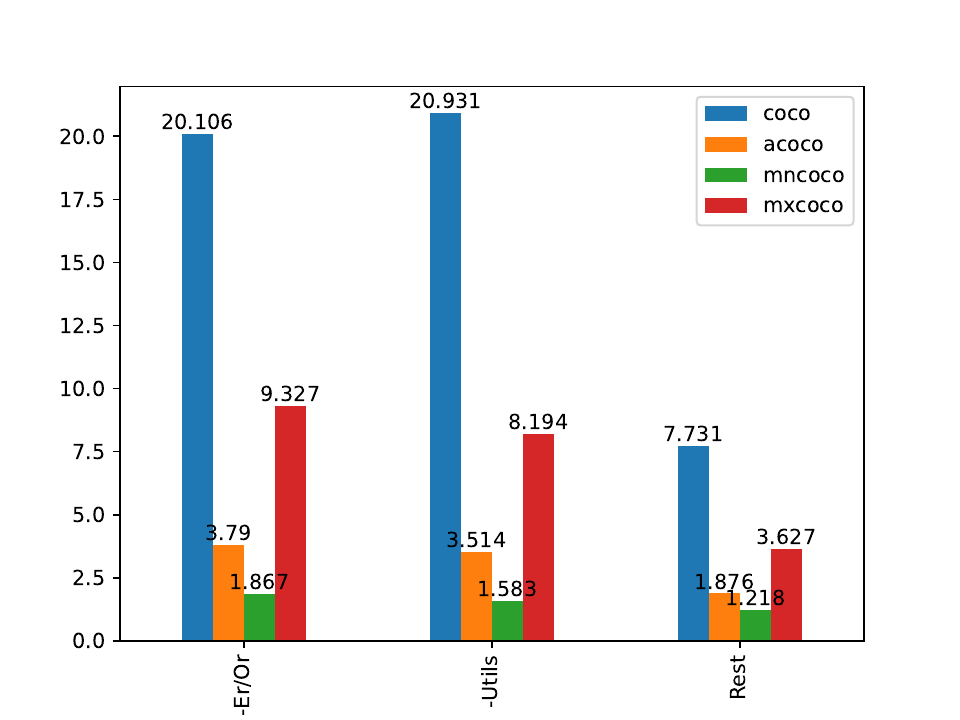}
\caption{The average values of total CoCo for three groups of Java classes: the classes with \eror{} and \utils{} suffixes are at least 2.5 times more complex.}
\label{fig:coco}
\end{figure}

\begin{figure}
\includegraphics[width=.9\columnwidth]{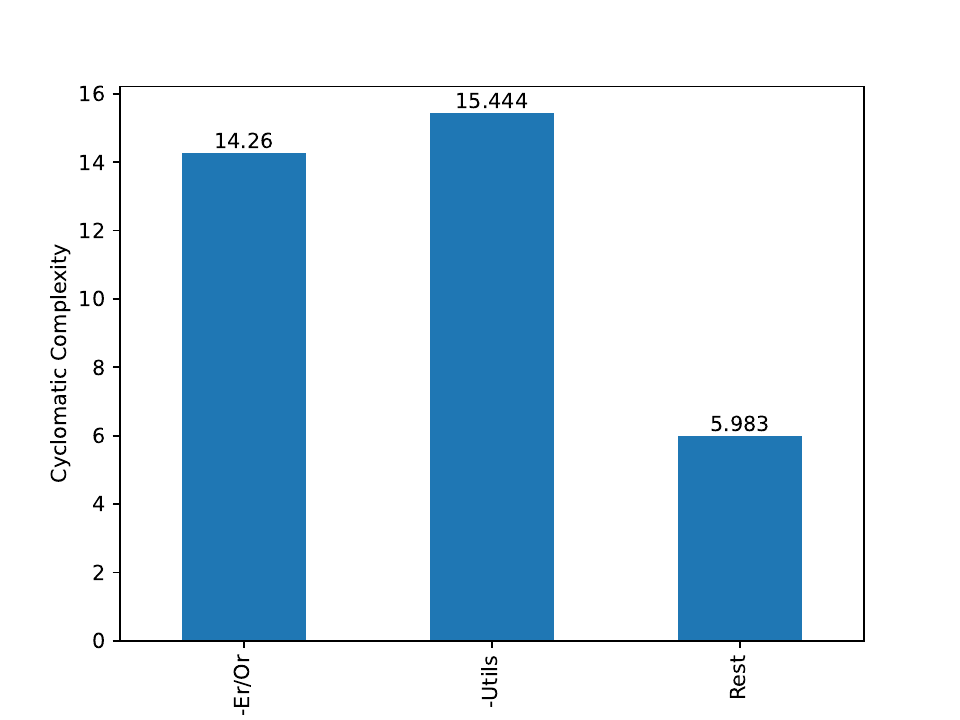}
\caption{The average values of total CC for three groups of Java classes: the classes with \eror{} and \utils{} suffixes are at least 2.5 times more complex.}
\label{fig:cc}
\end{figure}

Furthermore, \cref{fig:cc,fig:coco,fig:lcom,fig:nhd} illustrate separate bar charts for corresponding metrics.
It is visually obvious that the classes without \eror{} or \utils{} suffixes show the highest cohesion and the lowest complexity among all of the groups, except for the NHD metric. The classes with the \utils{} suffix have a slightly higher NHD value, thus better cohesion.

The entire dataset together with the scripts for its analysis is available at the open GitHub repository: \href{https://github.com/ILilliasI/sqm}{ILilliasI/sqm}.

\section{Discussion}
\label{sec:discussion}

\textbf{Why only LCOM5 and NHD were considered as cohesion metrics?}
In fact, there are many other metrics aimed at showing the degree of cohesion, complexity and other aspects of the code. It is best to include all the metrics possible, nevertheless, due to limited time we picked the described above as reflecting the core measurement of code analyzed in our research. It is also important to notice that all metrics were calculated exactly how they had been designed by authors.

\textbf{Is cohesion drop an indicator of proxy methods?}
The analysis of the collected data indicates that class names ending in `\eror{} and \utils{} are associated with a decrease in cohesion. This trend is evidenced by a decline in several key metrics. A deeper exploration into the causes  reveals that utility classes, which are prevalent in these scenarios, often employ an excess of proxy methods. These proxy methods are designed to simplify the interface for the users, allowing them to bypass certain arguments when calling a ``central'' utility method, as those arguments are set to default values. This approach, while seemingly convenient, may contribute to the overall reduction in cohesion.

Moreover, a common practice observed among developers handling \eror{} named classes is the aggregation of an extensive range of responsibilities within a single, bulky \ff{.java} file. This approach to class structuring poses significant challenges to maintainability. Such files can quickly become unwieldy and difficult to manage.

\textbf{How close functor classes are to ``God classes''?}
Indeed, an examination of \eror{} classes frequently uncovers the presence of what are colloquially known as ``God Classes.'' These are classes that are overly burdened with a multitude of responsibilities, attempting to oversee and execute an overly broad array of functions. This not only confuses the readers and users of such classes but also exacerbates the complexity of the codebase. Another issue is that they might induce is what is known as ``Fat Interface,'' where an interface offers more methods than clients require often mixing a variety of responsibilities similar to regular ``God classes,'' making the software difficult to extend or use.

\textbf{How exactly functors negatively affect quality of design?}
The presence of functors may act as an indicator of deeper structural and design issues within the code. Addressing these issues may not require only a reconsideration of naming practices but also an effort to adhere to principles of good software design, such as maintaining cohesion, ensuring modularity, and avoiding the pitfalls of creating overly complex classes that hinder the readability, maintainability, and extensibility of the code. In the context of utility classes, they typically encompass a ``central'' method that includes all essential arguments required to execute a particular task. This primary method is referred to by other similar methods, which supply default values to some arguments. This practice, however, tends to reduce cohesion and increase complexity.

\section{Limitations and Future Work}
\label{sec:limitations}

First, in our dataset there was no information about whether a class is \ff{abstract} or not. Because of this, we could not identify and exclude abstract classes as it had been done by \citet{10.1109/APSEC.2007.7}.

Next, in this work we considered metric values for a relatively small amount of utility classes due to a lack of sustainable data for those classes in dataset.

Additionally, to split all the classes in the dataset into considered groups, we made some assumptions. Primarily, we picked utility classes by criteria for a class name to end with \utils{} (and a few similar suffixes), but have not analyzed the underlying structure of a class to conclude whether it belongs to this group. The same applies to the \eror{} classes except for manual exclusion of all the classes which name ends in ``\ff{-User},'' ``\ff{-Server}'' as well as other suffixes. The full list can be found in \cref{sec:method}.

\section{Conclusion}
\label{sec:conclusion}

The main goal of our research was to find our whether functor Java classes have higher complexity and lower cohesion than all other classes. We took 13,861 Java classes from 212 open source repositories, divided them into three groups (\eror{} classes, \utils{} classes, and all other classes), and evaluated CC, CoCo, LCOM5, and NHD metrics in each group. Because average values of CC and CoCO in the first two groups were almost \emph{three times larger} that the values in the third group, we concluded that functor classes may be considered bad design decisions.

To further solidify the claims, more source code could be analyzed and more metrics could be calculated. Additionally, more sophisticated and robust classification algorithms for utility and \eror{} source code classes could be employed in future work. Moreover, several other object oriented programming languages could be observed, that is, for example, Python or C++.

\bibliographystyle{ACM-Reference-Format}
\bibliography{bibliography}

\end{document}